# Energy gain in a two-gap RF cavity

A. Shemyakin, Fermilab, Batavia, IL 60510, USA
*shemyakin@fnal.gov*

*Abstract*
Velocity of a deeply non-relativistic particle can change during its acceleration inside an RF cavity significantly enough to cause a deviation of the energy gain from a linear model. This paper derives formulae for corrections to the energy gain in a two-gap RF cavity calculated by taking into account this velocity change in the next approximation. Then, the results are compared with direct particle tracking of an HWR cavity used at the PIP2IT test accelerator. The formulae significantly extend the range of parameters where analytical calculations work with a good accuracy. The approach is applied to two aspects of cavity phasing.

## Table of Contents

## 1. Introduction

The work presented in this paper is related to attempts to understand peculiarities of acceleration in cavities of the test accelerator PIP2IT [1], which models the front end of the future PIP-II linac [2]. In PIP2IT, the $H^-$ beam is generated in an ion source, travels through the LEBT, is accelerated in the RFQ to 2.1 MeV, transported through the MEBT, and then accelerated by individually controlled SRF cavities in two cryomodules, Half-Wave Resonator (HWR) and Single-Spoke Resonator (SSR1). The high rate of acceleration results in deviation of the energy gain in a cavity from a linear model to the level already important for operation.

All RF cavities used in PIP2IT (apart of RFQ) have a similar two-gap structure. Electric field in such cavity can be approximated well by a sine function. It allows to derive formulae for the first non-linear correction to the energy gain and compare them with particle tracking. These analytical expressions allow to see more clearly deviations from the linear model and to make more accurately simple estimations without computer simulations.

Section 2 presents the energy gain formulae for a two-gap cavity in the linear approximation following Ref. [3] and repeating the derivation made (by other authors) for PIP-II CDR [4]. Section 3 extends the analytical description to the second order. Sections 4 and 5 compare the formulae with direct tracking through electric field of a pure sinusoid and of a real cavity, correspondingly. Section 6 considers consequences of the non-linear behavior for practical cavity phasing.

All numerical examples in this paper are shown for the HWR cavity [5].

## 2. Sinusoidal field, first order

Let's consider a particle with charge $e$ and initial kinetic energy $W_0$ passing through an RF cavity which electric field on axis is

$$E_z(z,t) = \begin{cases} E_0 \sin(k_s z)\sin(\omega t + \varphi_0), -\frac{L}{2} \le z \le \frac{L}{2}, \\ 0, otherwise \end{cases} \tag{1}$$

where $k_s = \frac{2\pi}{L}$, and time is counted from the moment when the particle is at $z = 0$. After passing the cavity, the particle gains the energy

$$\Delta W = \int_{-\frac{L}{2}}^{\frac{L}{2}} eE_z\big(z, t(z)\big)\, dz \tag{2}$$

In the first approximation, the velocity change inside the cavity is neglected, and

$$\omega t(z) \approx \frac{\omega z}{\beta_0 c} \equiv kz, \tag{3}$$

where $\beta_0$ is the relativistic factor at energy $W_0$, and $c$ is the speed of light. The integral in Eq.(2) is expressed as

$$\Delta W_1 = \int_{-\frac{L}{2}}^{\frac{L}{2}} eE_0 \sin(k_s z)\sin(kz + \varphi_0)\, dz = \frac{eE_0 L}{2\pi}\int_{-\pi}^{\pi} \sin\varphi \sin(\alpha\varphi + \varphi_0) d\varphi =$$

$$= \frac{eE_0 L}{2\pi}\left[\frac{\sin\big(\pi(1-\alpha)\big)}{1-\alpha} - \frac{\sin\big(\pi(1+\alpha)\big)}{1+\alpha}\right]\cos\varphi_0, \tag{4}$$

$$\varphi \equiv k_s z\,;\ \alpha \equiv \frac{k}{k_s}$$

where variables $\varphi$ and $\alpha$ are introduced to simplify the expression. It is convenient to express the result of Eq.(4) as a product of three components

$$\Delta W_1 = eV_g \cdot T(\alpha) \cdot \cos\varphi_0, \tag{5}$$

The first component, the cavity voltage $eV_g$, defines the dependence on the cavity field amplitude

$$eV_g = \frac{eE_0 L}{2}. \tag{6}$$

Numerically, it is the maximum energy gain at the initial velocity corresponding to the "geometrical beta"

$$\beta_G = \frac{\omega L}{2\pi c}. \tag{7}$$

Note that with this definition the variable α introduced in Eq.(4) can be expressed as

$$\alpha = \frac{\beta_G}{\beta_0}. \tag{8}$$

The second component, which defines the dependence of the energy gain, is the transit time factor $T(\alpha)$ normalized to 1 at the geometrical beta (at $\alpha = 1$). It can be simplified to

$$T(\alpha) = \begin{cases} \frac{2\sin\pi\alpha}{\pi(1-\alpha^2)}, \alpha \neq 1 \\ 1, \qquad \alpha = 1 \end{cases}. \tag{9}$$

The last factor in Eq.(5) is the cosine dependence on the cavity phase, which is defined by Eq.(1) and Eq. (3) for the moment when the particle crosses the central plane of the cavity.

Fig. 1 shows the function $T(\alpha)$. It reaches maximum $T_{opt} = 1.042$ at the velocity $\beta_{opt}c$ slightly higher than $\beta_G c$, corresponding to $\alpha_{opt} = \frac{\beta_G}{\beta_{\text{opt}}} = 0.837$.

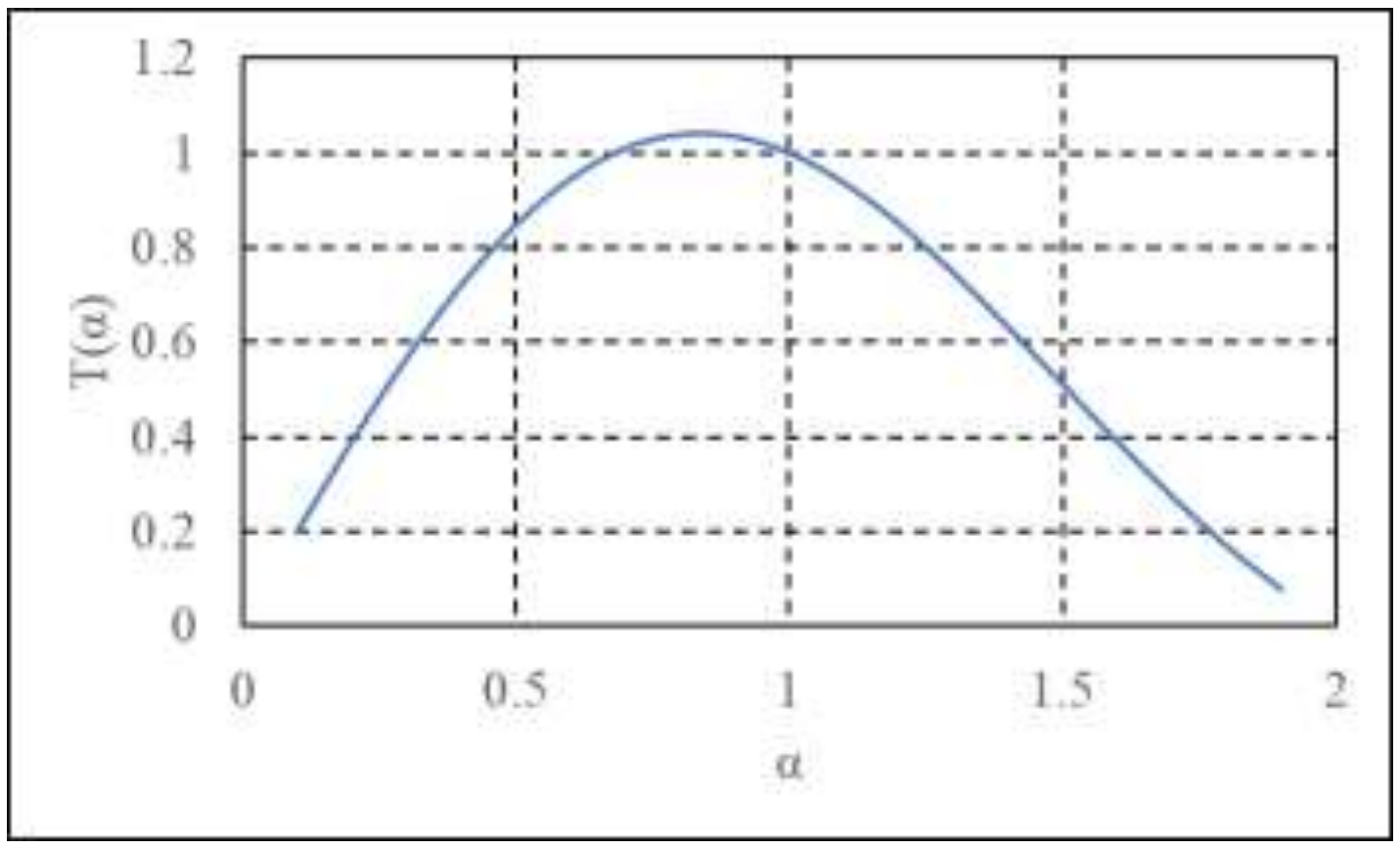

Figure 1. Transit time factor dependence on the initial velocity expressed as $\alpha_{opt} = \frac{\beta_G}{\beta_{\text{opt}}}$.

In this paper, the term "cavity voltage" is used for the maximum energy gain at the geometrical beta since this value naturally appears in formulae. Note, however, that this definition differs from the one used in PIP2IT simulations and operation, where "cavity voltage" refers to the maximum gain at the optimum beta.

## 3. Sine function, second order

The accuracy of prediction of the energy gain in a sine-function electric field can be improved by taking into account to the first order the energy change inside the cavity. In this approximation, Eq.(3) is modified into

$$\omega t(z) = \frac{\omega}{\beta_0 c} \int_{-\frac{L}{2}}^{z} \frac{\beta_0}{\beta(z)} dz - \alpha\pi, \tag{10}$$

The second term on the right hand side of Eq.(10) preserves the definition of the cavity phase used in the previous section for the case of a low cavity voltage. Changes in velocity can be estimated from the energy gain $\delta W(z)$ calculated in first approximation:

$$\frac{\beta_0}{\beta(z)} \approx 1 - \frac{\delta\beta(z)}{\beta_0} \approx 1 - \frac{\delta W(z)}{W_0\gamma_0(\gamma_0+1)} \approx$$

$$\approx 1 - \frac{1}{W_0\gamma_0(\gamma_0+1)} \int_{-\frac{L}{2}}^{z} eE_0 \sin(k_s z_1)\sin(kz_1+\varphi_0)\, dz_1, \tag{11}$$

where $\gamma_0$ is the relativistic factor at the initial energy. Using Eq.(11) and notations of Eq.(4), Eq.(10) can be written as

$$\omega t(\varphi) \approx \alpha\varphi - \alpha\cdot\delta\varphi(\varphi),$$

$$\delta\varphi(\varphi) \equiv \frac{1}{W_0\gamma_0(\gamma_0+1)} \frac{eE_0L}{2\pi} \int_{-\pi}^{\varphi} d\varphi_1 \int_{-\pi}^{\varphi_1} \sin\varphi_2 \sin(\alpha\varphi_2+\varphi_0) d\varphi_2\,. \tag{12}$$

The expression for the total energy gain is linearized assuming that $\delta\varphi \ll 1$:

$$\Delta W \approx \frac{eE_0L}{2\pi} \int_{-\pi}^{\pi} \sin\varphi \sin(\alpha\varphi - \alpha\cdot\delta\varphi(\varphi)+\varphi_0)\, d\varphi \approx$$

$$\approx \frac{eE_0L}{2\pi}\left[\int_{-\pi}^{\pi} \sin\varphi\sin(\alpha\varphi+\varphi_0)\, d\varphi - \int_{-\pi}^{\pi} \sin\varphi\cos(\alpha\varphi+\varphi_0)\cdot\alpha\cdot\delta\varphi(\varphi)\, d\varphi\right] \equiv$$

$$\equiv \Delta W_1 - \Delta W_2. \tag{13}$$

Combining Eq.(12) and Eq.(13) gives the correction term $\Delta W_2$ as

$$\Delta W_2 = \frac{1}{W_0\gamma_0(\gamma_0+1)}\left(\frac{eE_0L}{2\pi}\right)^2 \alpha\,\cdot$$

$$\cdot \int_{-\pi}^{\pi} \sin\varphi\cos(\alpha\varphi+\varphi_0)\, d\varphi \int_{-\pi}^{\varphi} d\varphi_1 \int_{-\pi}^{\varphi_1} \sin\varphi_2\sin(\alpha\varphi_2+\varphi_0) d\varphi_2\,. \tag{14}$$

The integral in Eq.(14) can be taken analytically and is equal to

$$\frac{1}{4\alpha(1-\alpha^2)^3}\cdot\left\{\begin{matrix} 8\alpha^2(1-\cos 2\pi\alpha) + 4\pi\alpha(1-\alpha^2)\sin 2\pi\alpha - \\ -\sin 2\varphi_0\,[(1+3\alpha^2)\sin 2\pi\alpha + 4\pi\alpha(1-\alpha^2)] \end{matrix}\right\}. \tag{15}$$

For convenience, the complete expression for the cavity energy gain in the second order is shown below with repeating of definitions of the main variables.

$$\Delta W \approx eV_g\cdot T(\alpha)\cdot\cos\varphi_0 + \frac{(eV_g)^2}{W_0\gamma_0(\gamma_0+1)}[T_1(\alpha) + T_2(\alpha)\sin 2\varphi_0],$$

$$T(\alpha) = \begin{cases} \dfrac{2\sin\pi\alpha}{\pi(1-\alpha^2)}, \alpha \neq 1 \\ 1, \qquad \alpha = 1 \end{cases},$$

$$T_1(\alpha) = \begin{cases} -\dfrac{8\alpha^2(1-\cos 2\pi\alpha) + 4\pi\alpha(1-\alpha^2)\sin 2\pi\alpha}{4\pi^2(1-\alpha^2)^3}, \alpha \neq 1 \\ \dfrac{1}{4}, \qquad \alpha = 1 \end{cases}$$

$$T_2(\alpha) = \begin{cases} \dfrac{(1+3\alpha^2)\sin 2\pi\alpha + 4\pi\alpha(1-\alpha^2)}{4\pi^2(1-\alpha^2)^3}, \alpha \neq 1 \\ \dfrac{8\pi^2-3}{48\pi}, \qquad \alpha = 1 \end{cases}$$

$$\beta_G = \frac{\omega L}{2\pi c};\ \alpha = \frac{\beta_G}{\beta_0};\ eV_g = \frac{eE_0 L}{2}. \qquad (16)$$

The functions $T_1(\alpha)$ and $T_2(\alpha)$ are shown in Fig.2.

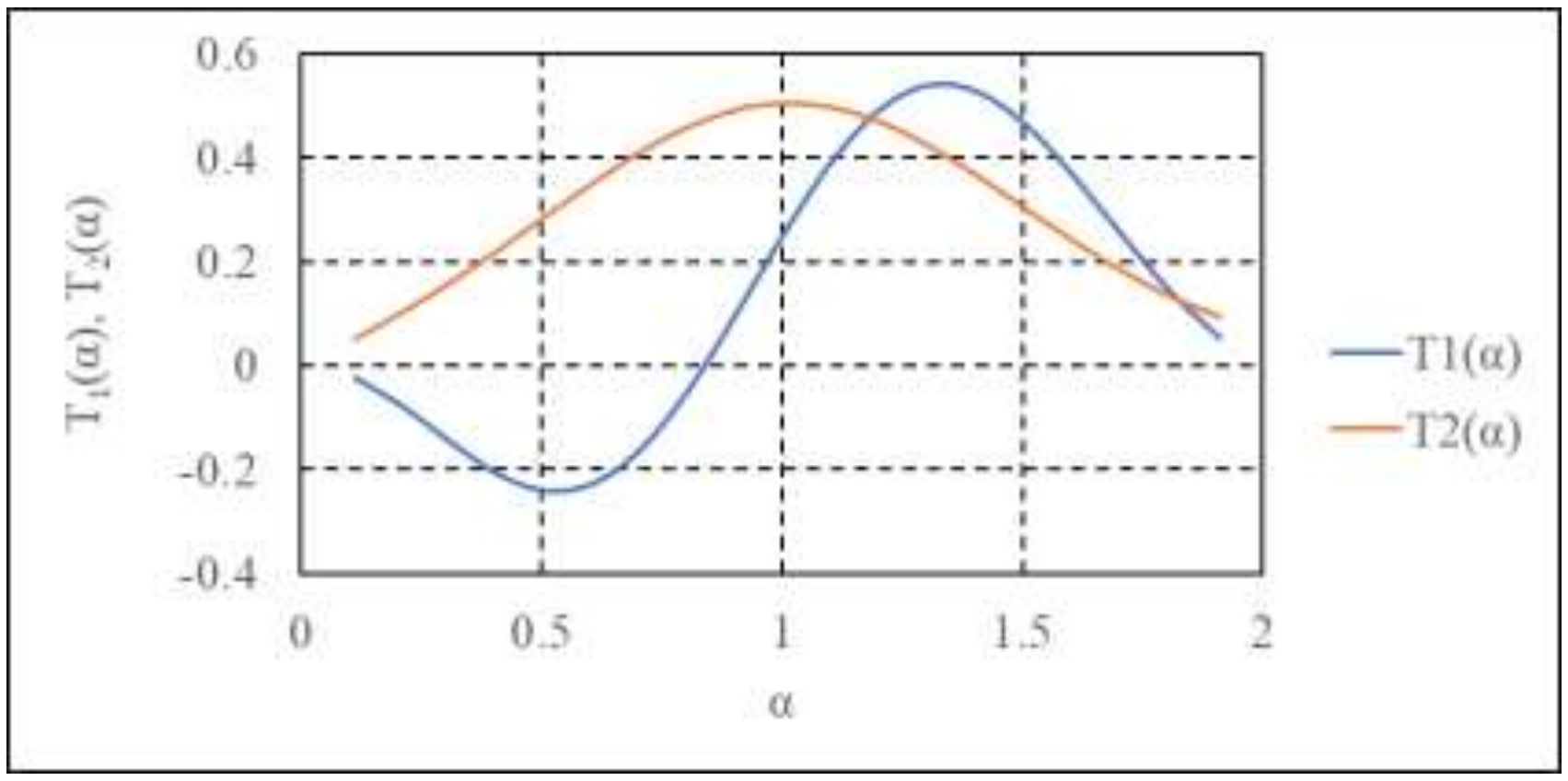

Figure 2. Correction functions for the energy gain.

At the optimum energy, $T_1(\alpha_{opt})$ is close to zero. Since the second component of the correction is zero at $\varphi_0 = \frac{\pi}{2}n$, the energy gain at optimum beta stays zero at the same phase ($\varphi_0 = \pm\frac{\pi}{2}$) independently on the cavity voltage (within the model assumption of small amplitudes, $eV_g \ll W_0$). At other energies, a zero-energy gain does not correspond anymore to $\varphi_0 = \pm\frac{\pi}{2}$.

In this approximation, the phase providing the maximal gain deviates from zero, and the difference between phases with the maximum gain (usually accepted as "zero phase") and with zero gain ("bunching phase") is not exactly 90º.

## 4. Comparison with direct tracking in the case of the sine function

Calculation with Eq.(16) was compared with tracking of particle ($H^-$) motion in the field of Eq.(1). Tracking is made in MathCad with a built-in function that uses the fourth-order Runge-Kutta method with adaptive step-size. Some of the results are shown in Fig. 3. Also, the case when the energy gain is calculated in the first approximation with Eq.(4) is presented in the right column.

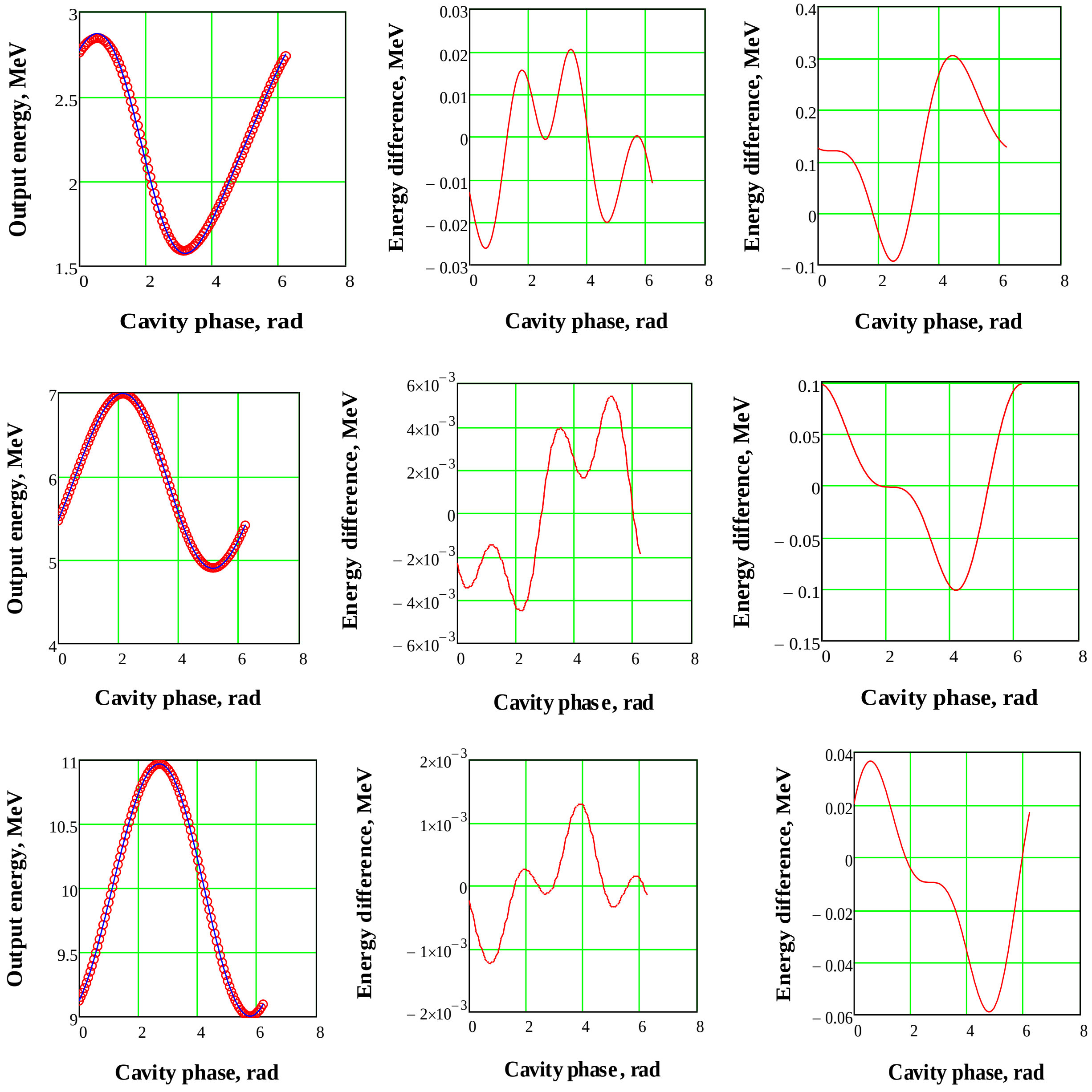

Figure 3. Comparison of calculation with Eq.(16) and results of direct tracking. Cavity length $L$ = 0.173 m, corresponding to $\beta_{opt}$ =0.112 ($W_{opt}$=5.9 MeV for $H^-$), which represents the PIP2IT HWR cavity. $H^-$ ion is injected with ), 2.1 MeV (top row), $W_0 = W_{opt}$ (middle row), and 10 MeV (bottom row). Cavity voltage $V_g$ =1 MV. Left column: energy gain calculated with Eq.(16) is shown as the blue curve, and direct integration is represented by red circles. Center: difference between two curves of the left plot. Right: the difference between tracking and using the first approximation only (Eq.(4)).

The deviation of the models from the tracking results (curves in the central and right columns of Fig.3) depends dramatically on the choice of the phase offset. In Fig.3, the phase for the tracking results is defined for the moment when the particle is at the beginning of the field region (i.e. at $z = -\frac{L}{2}$). Phases for the curves representing Eq.(16) and Eq.(4) are adjusted to have the maximum energy gain at the same point as for tracking. If, for example, the phase is chosen by the best fitting of cosine to the entire tracking curve, the difference decreases by several times in comparison with the right column and is defined mainly by the second harmonic.

Eq.(16) describes the energy gain well for the practically interesting range, 2-10 MeV for the modelled PIP2IT HWR cavity with $W_{opt}$=5.9 MeV at the cavity voltage $V_g$ =1 MV. The error increases for lower energies, but even for $W_0 = 2.1$ MeV (top row), when $eV_g$ is comparable with $W_0$, the energy gain is predicted with several percent accuracy. For this set of parameters, using only the first approximation gives an error by an order of magnitude larger.

The phase of the maximum gain changes with the cavity voltage. In the considered approximation, the phase shift can be found by derivation of the energy gain over the phase in Eq.(16). The derivative is equal zero at the phase $\varphi_{0m}$:

$$\sin\ \varphi_{0m} = \frac{1}{4 \cdot A_1}\left(\sqrt{1 + 8 \cdot A_1^2} - 1\right), A_1 \equiv \frac{2 \cdot eV_g}{W_0\gamma_0(\gamma_0 + 1)} \cdot \frac{T_2(\alpha)}{T(\alpha)}. \tag{17}$$

This shift compares well with what is found in tracking for the initial phase corresponding to the maximum gain (Fig. 4). At the low ratios of $\frac{eV_g}{W_0\gamma_0(\gamma_0+1)}$ the shift is close to the change in the arrival time to the cavity center.

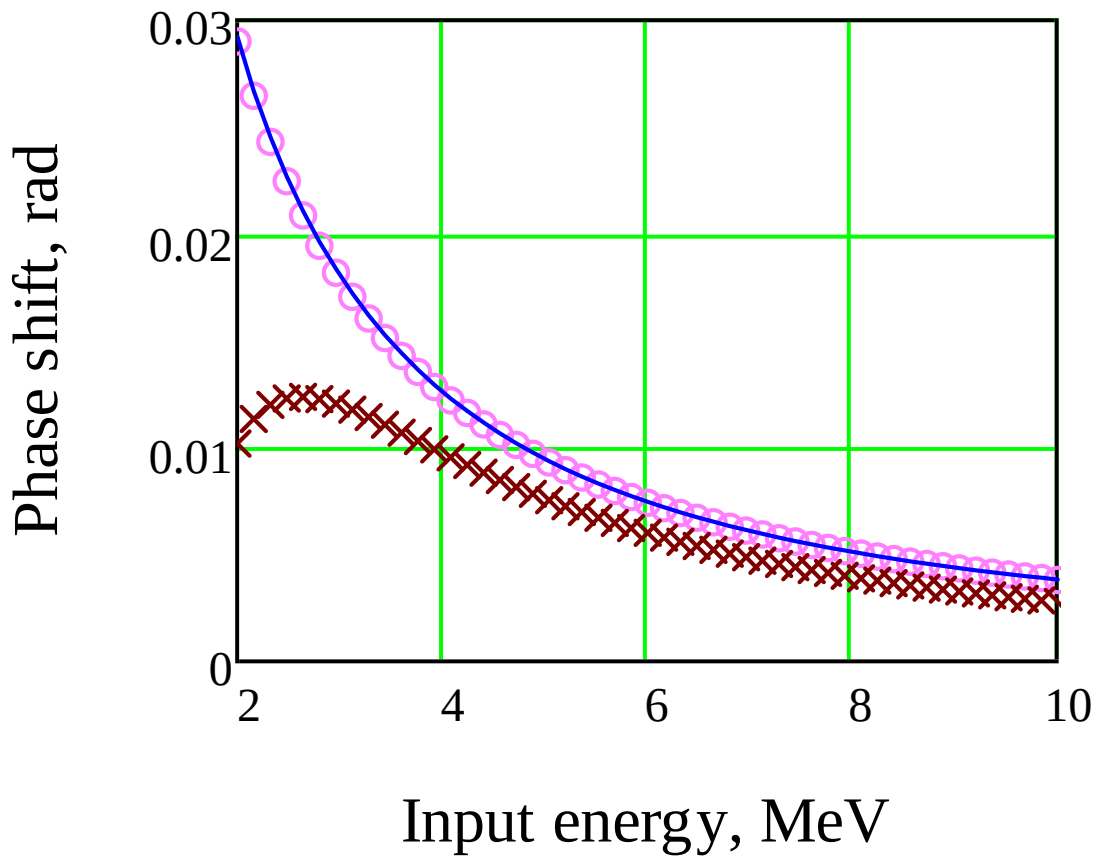

Figure 4. Cavity phase corresponding to the maximum gain vs initial energy. The cavity voltage is $V_g = 1$ MV. Pink circles correspond to the changes in the optimum initial phase in tracking (after subtracting the flight time to the center of the cavity $\omega\frac{L}{2\beta_0 c}$). The blue curve is calculated with Eq.(12). The crosses show deviation of the time of arrival to the cavity center in tracking.

Note that the maximum energy gain at $\beta_{opt}$ stays close to the one defined by Eq.(4). For example, for $V_g$= 2 MV, $W_0$=6 MeV, the deviation is 1%. Hence, the operational definition of the "cavity voltage" as the maximum gain at $\beta_{opt}$ stays well in the entire range of PIP2IT parameters.

## 5. Comparison with direct tracking through a field map

The field of a real cavity differs from approximation of Eq. (1). As an example, Fig. 5 shows the field distribution in the HWR cavity simulated for the actual cavity geometry [6].

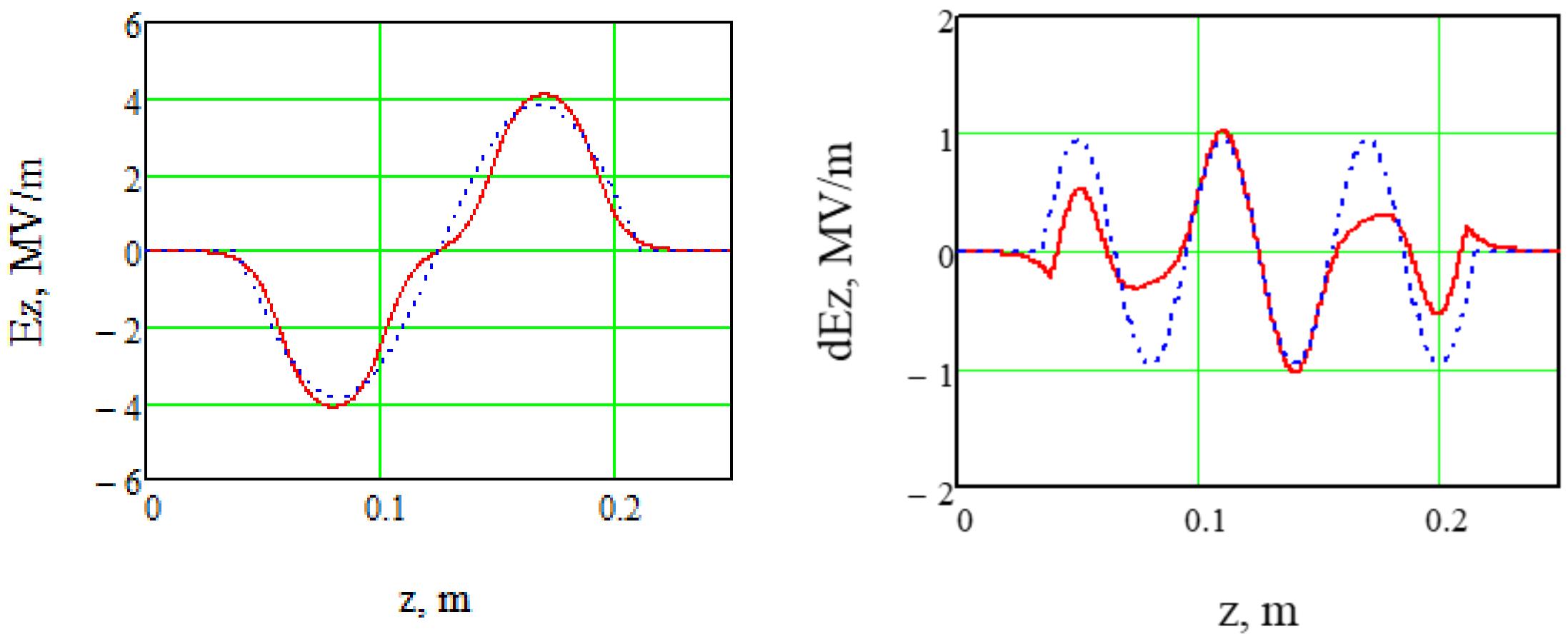

Figure 5. Left: Field map of HWR cavity (red) and fit by sine function (blue). Right: difference between two curves on left (red). The blue curve is a sine function of triple spatial frequency to guide the eye. The field map corresponds to $eV_g$ =0.316 MeV.

There is an uncertainty in defining a priori parameters of the sine function in Eq.(1) to optimally reproduce the cavity performance. In can be done using the results of particle tracking in the following steps.

1. With the cavity phase set to provide the maximum gain, scan the input energy with the cavity voltage constant and corresponding to a small cavity voltage ($eV_g \ll W_0$). Find the energy $W_{opt}$ where the gain is maximal, $\Delta W_{max}$. Calculate the corresponding $\beta_{opt}$.
2. Calculate the geometrical beta based on the sine model as $\beta_G = \alpha_{opt}\beta_{\mathrm{opt}} = 0.837 \cdot \beta_{\mathrm{opt}}$. Calculate the effective cavity length $L = \frac{2\pi\beta_G c}{\omega}$.
3. Calculate the effective cavity voltage for the given electric field amplitude as $eV_g = \frac{\Delta W_{max}}{T_{opt}} = \frac{\Delta W_{max}}{1.042}$.

The sine function determined in this manner for the HWR field map is shown as the blue curve in Fig. 5. Fig. 6 compares the maximal energy gains produced by tracking in MathCad through the actual field represented by the red curve of Fig. 5 and by calculation with Eq.(16) using parameters for the fitting sinusoid.

Two energy gain curves in Fig. 6 are visually significantly closer than two field distributions in Fig. 5 generating them. To interpret that, one can note that in the linear approximation of Eq.(4), which is valid at low $eV_g$, the energy gain can be presented as a sum of the gains from the sine distribution and from the difference between the actual and sine distributions. The difference, presented in Fig. 5 right, has ~3 times higher spatial frequency than the main component and, correspondingly, can contribute to the final energy gain significantly only the at low input energies (correspondingly, by ~9 times lower than $W_{opt}$, or ~0.6 MeV).

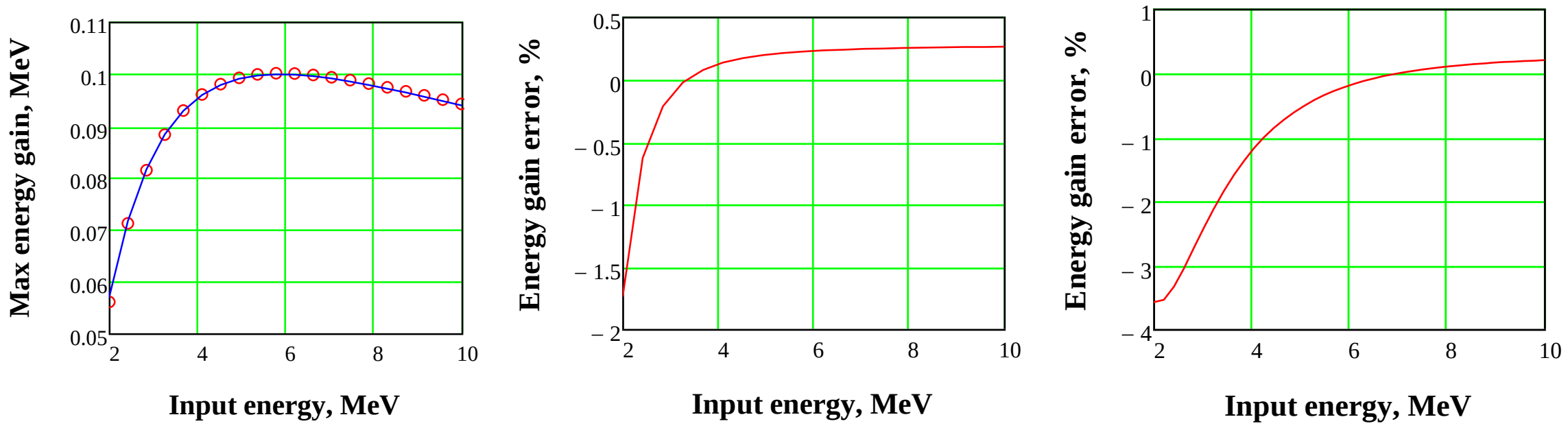

Figure 6. Left: the maximum energy gain simulated with tracking through the HWR field map (red circles) and calculated with Eq.(16) (blue curve) vs input energy at $eV_g = \frac{0.1\ MeV}{1.042}$. Center: the relative difference between the two curves on the left. Right: the same relative difference for the case of $eV_g = \frac{1\ MeV}{1.042}$.

## 6. Notes on cavity phasing

The analytical formulae presented in the previous sections describe well the energy gain in a useful range of parameters. This approach helps to understand better peculiarities of the cavity phasing as it was implemented at PIP2IT.

In PIP2IT operation, the phase of a cavity is defined as zero when the beam energy gain is maximal. To find the offset in low-level RF system corresponding to this relationship, the cavity phase is varied while the beam phases in the downstream BPMs are recorded. Let's assume that the absolute phases of the BPMs are calibrated and stable. In this case, the measured difference between two BPMs phases provides the beam velocity, and, therefore, a set of energy values vs cavity phases.

The first peculiarity is related to phasing in the case when the condition $eV_g \ll W_0$ is not satisfied. To find the cavity phase corresponding to the maximum energy in the phase scan data, one needs to fit a curve to this set. The data interval should cover enough data points deviating from the maximum by much more than the BPM measurement noise. If the fitting curve is a cosine, the result can deviate significantly from the true position and depends on the choice of the interval because of the asymmetry of the actual energy gain vs cavity phase. This asymmetry, signified in Eq.(16) by the $\sin 2\varphi_0$ term, becomes important for the large relative energy gains.

As an example, Fig. 7 emulates a phase scan of a HWR cavity with the entry energy of 2.1 MeV and the cavity voltage corresponding to the energy gain at $\beta_{opt}$ of 0.6 MeV. The energy after the cavity is simulated by particle tracking through the cavity field map and fitted to cosine. If the phase scan interval covers 360º, the error in the fitted position of maximum is 9º. In this simulated case with no measurement errors, the fit accuracy improves for smaller intervals. If instead the fit uses Eq.(16), the result does not depend on the interval within <1º.

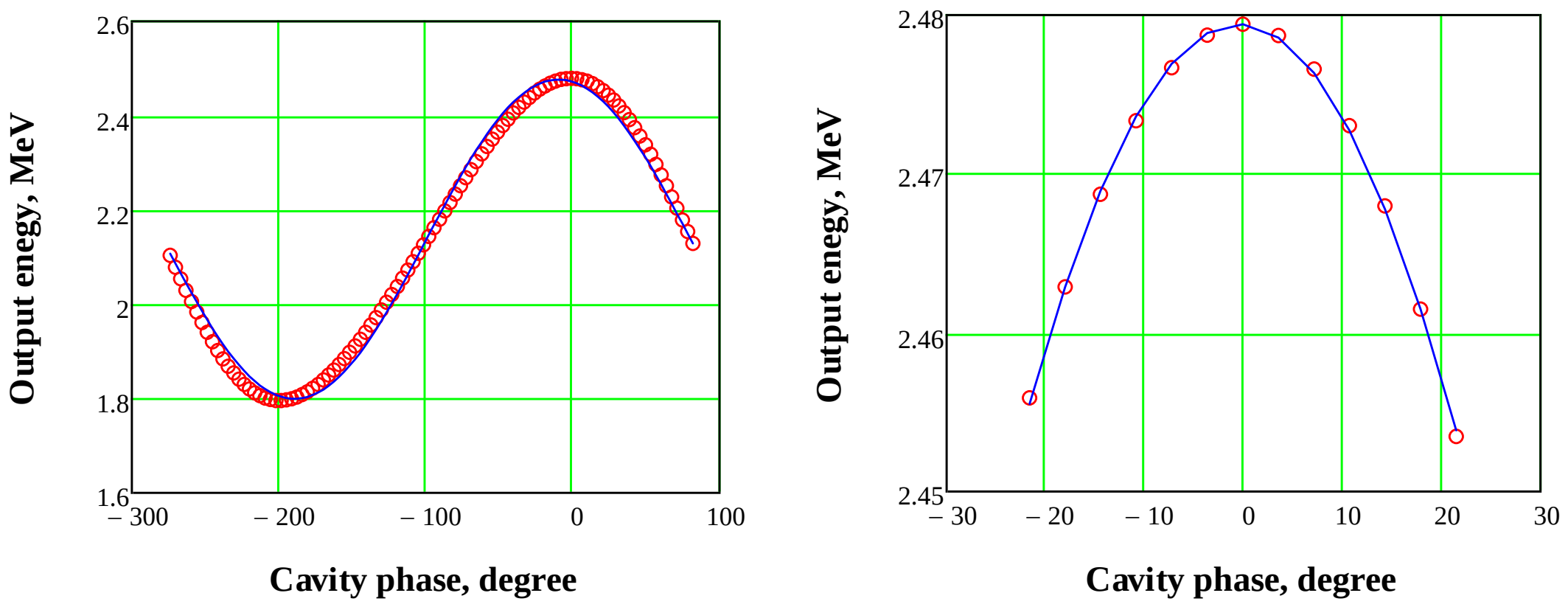

Figure 7. Fitting of the cosine (blue) to "phasing curve" data simulated by particle tracking (red) for the case of two intervals. The fitted position of the maximum differs between these two cases by 8.7º. The entry energy of is 2.1 MeV, and $eV_g = \frac{0.6\ MeV}{1.042}$.

The second note is about a possibility to use only one BPM for cavity phasing, which was attempted at some stage of PIP2IT commissioning. It is simpler in operation because such procedure minimizes the number of downstream cavities that need to be turned off or, as it was in the case of the HWR cryomodule, eliminates this need completely. The approach used in Section 3 can be used to look at the phasing error caused by this procedure.

One-BPM phasing is valid if the effect of the cavity can be approximated well by the energy jump localized in its central plane so that the total phase change over its length is

$$\psi_c = \frac{\omega}{c}\left(\frac{z_c - z_0}{\beta_0} + \frac{z_f - z_c}{\beta_f}\right), \tag{18}$$

where $z_0$, $z_c$, and $z_f$ are the coordinates of the field map start, center, and end; and $\beta_f$ is the relativistic factor calculated with the final energy. In this case, minimum of the BPM phase reading corresponds to the maximum energy gain, and, therefore, to zero cavity phase as defined above.

Applicability of such approach in practice can be estimated using the sinusoidal field model. In this case, the total phase change over the cavity length starting at $z_0 = -\frac{L}{2}$ is given by Eq.(12) evaluated at the end of the cavity:

$$\psi_{mod} = \omega\left[t\left(\frac{L}{2}\right) - t\left(-\frac{L}{2}\right)\right] \approx 2\pi\alpha - \alpha\cdot\delta\varphi(\pi),$$

$$\delta\varphi(\pi) \equiv \frac{1}{W_0\gamma_0(\gamma_0+1)}\frac{eE_0L}{2\pi}\int_{-\pi}^{\pi} d\varphi_1 \int_{-\pi}^{\varphi_1} \sin\varphi_2 \sin(\alpha\varphi_2 + \varphi_0)d\varphi_2\ . \tag{19}$$

Taking the integral yields

$$\int_{-\pi}^{\pi} d\varphi_1 \int_{-\pi}^{\varphi_1} \sin\varphi_2 \sin(\alpha\varphi_2+\varphi_0)d\varphi_2 =$$

$$= \begin{cases} \pi^2 T(\alpha)\cos\varphi_0 - \dfrac{2\pi}{1-\alpha^2}[\alpha T(\alpha) + \cos 2\pi\alpha]\sin\varphi_0\,, \alpha \neq 1 \\ \dfrac{\pi}{2}\sin\varphi_0 + \pi^2\cos\varphi_0\,, \qquad \alpha = 1 \end{cases} \tag{20}$$

Using Eq.(20) and notation of Eq.(16), Eq. (19) is transformed into

$$\psi_{mod} = 2\pi\alpha\left(1-\frac{\Delta W_1}{2W_0\gamma_0(\gamma_0+1)}\right)+\frac{eV_g}{W_0\gamma_0(\gamma_0+1)}\frac{2\alpha}{1-\alpha^2}[\alpha T(\alpha)+\cos 2\pi\alpha]\sin\varphi_0 \quad (21)$$

The first term of Eq.(21) describes the model of Eq.(18). The second term has the same dependence on the cavity voltage and energy as the one in brackets, which means that, generally speaking, the phasing procedure with a single BPM is not valid independently on the cavity voltage.

In practice, however, the changes in the beam phase are measured in the BPM separated by a distance $S_{BPM}$ from the cavity center, and typically $S_{BPM} \gg L$. The particle travels the length outside of the cavity $S_{BPM} - {}^{L}/_{2}$ with the speed of $\beta_f c$, and the phase measured by the BPM (counted by the time of the particle entering the cavity at $z_0 = -{}^{L}/_{2}$) is

$$\psi_{BPM} = \psi_{mod} + \omega\frac{S_{BPM} - {}^{L}/_{2}}{\beta_f c} = \psi_{mod} + \pi\alpha\left(\frac{2S_{BPM}}{L}-1\right)\frac{\beta_0}{\beta_f}. \quad (22)$$

Using the same approximation as in Eq.(11) and ignoring second-order corrections to the energy gain,

$$\frac{\beta_0}{\beta_f} \approx 1 - \frac{\Delta W_1}{W_0\gamma_0(\gamma_0+1)}. \quad (23)$$

Combining Eq.(21), (22), (23), the final expression is

$$\psi_{BPM} = \pi\alpha\left(\frac{2S_{BPM}}{L}+1\right) - \pi\alpha\frac{\Delta W_1}{W_0\gamma_0(\gamma_0+1)}\frac{2S_{BPM}}{L} +$$
$$+\frac{eV_g}{W_0\gamma_0(\gamma_0+1)}\frac{2\alpha}{1-\alpha^2}[\alpha T(\alpha)+\cos 2\pi\alpha]\sin\varphi_0\,. \quad (24)$$

The first term in the sum of Eq.(24) corresponds to the flight time with the initial velocity independent on the cavity settings; the second one represents the change of the flight time from the cavity center to the BPM caused by the gained energy; and the last one is the phase change inside the cavity related to the second-order correction.

At low cavity voltage, when Eq.(5) is accurate, the maximal energy gain is at $\varphi_0 = 0$, while the minimum of the BPM phase predicted by Eq.(24) is described by the function

$$\tan\varphi_{min}(\alpha) = T_{corr}\frac{L}{2S_{BPM}},$$
$$T_{corr}(\alpha) = \frac{2}{\pi(1-\alpha^2)}\frac{\alpha T(\alpha)+\cos 2\pi\alpha}{T(\alpha)}. \quad (25)$$

The function $T_{corr}(\alpha)$ is shown in Fig. 8, left for the range relevant for PIP2IT ($W_0$ =2.1 – 10 MeV for HWR cavities). The deviation is small in a narrow vicinity of $\alpha_{opt}$=0.837. Fig. 8, right shows the error of phasing calculated with Eq.(25) for the ratio of $\frac{L}{2S_{BPM}}$ =0.179 representing parameters in the HWR cryomodule. Phasing with a single BPM provides error below 1° in the range of 4.6 – 7.5 MeV.

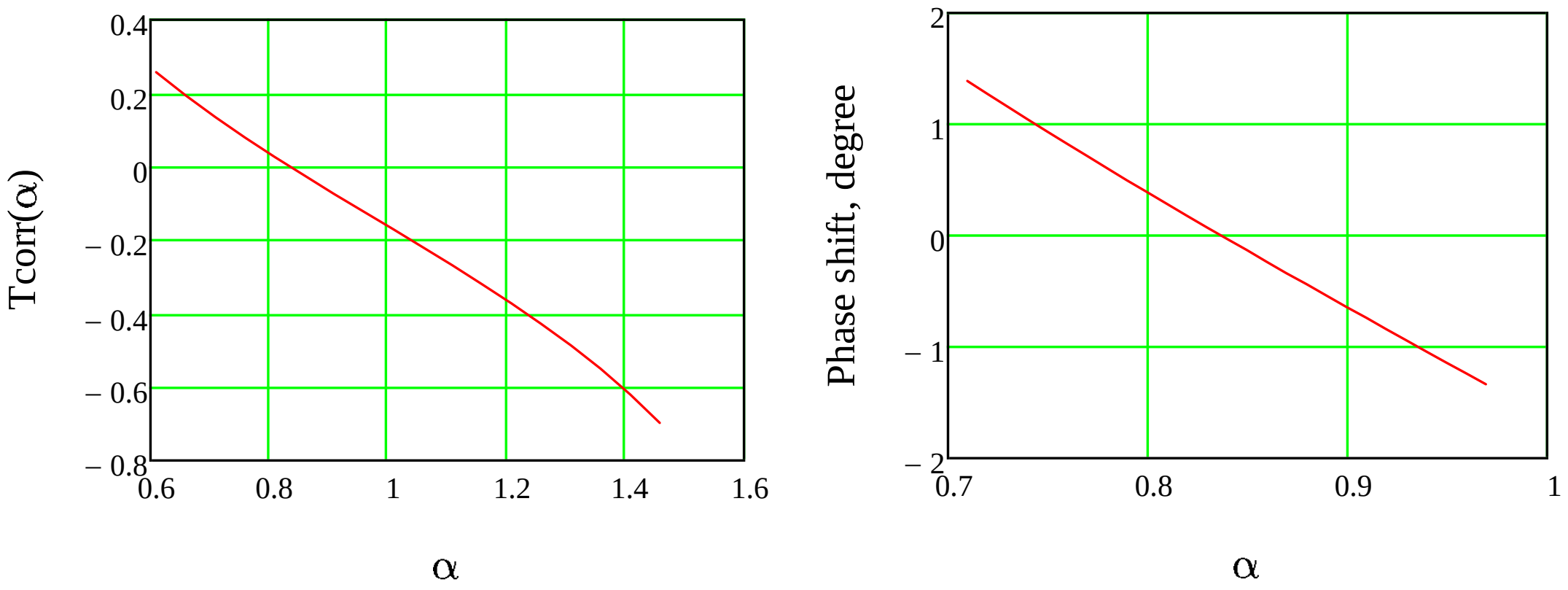

Figure 8. Difference between the cavity phases providing the maximum energy gain and minimum phase in a downstream BPM as a function of the particle energy expressed as $\alpha = \frac{\beta_G}{\beta_0}$. Left - function $T_{corr}(\alpha)$. Right- example of the difference in vicinity of $\alpha_{opt}$ , in degrees, for the distances of the HWR cryomodule.

## 7. Conclusion

Particle acceleration in two-gap cavities used at PIP2IT is represented well by the model employing a sinusoidal distribution of the longitudinal electric field. Second-order correction to the energy gain in a cavity is calculated by taking into account changes of particle velocity inside the cavity. The resulting formulae significantly extends the range of parameters where analytical calculations work with a good accuracy.

At PIP2IT parameters, the energy gain can significantly deviate from the cosine dependence on the cavity phase, complicating the phasing procedure. Possibility to use for phasing the second-order formulae can be considered.

The cavities can be phased using a single BPM if the beam energy is not too far from the optimum for the cavity.

## 8. Acknowledgement

This manuscript has been authored by Fermi Research Alliance, LLC under Contract No. DE-AC02-07CH11359 with the U.S. Department of Energy, Office of Science, Office of High Energy Physics.